\documentclass[cameraready]{Interspeech}

\title{Dialogs: a studio-quality expressive conversational Russian speech corpus for dialog assistants}

\author[affiliation={1}, orcid=0000-0002-1951-1113, equalcontribution]{Ilya}{Shigabeev}
\author[affiliation={1}, orcid=0009-0004-2485-485X, equalcontribution]{Ilya}{Latyshev}

\address{
    $^1$ Langswap, Russia
}

\email{shigabeevilya@gmail.com, sokolilia56@gmail.com}


\usepackage{comment}
\usepackage{amssymb}

\begin{document}

\maketitle

\begin{abstract}
    We introduce Dialogs, a studio-quality Russian conversational speech corpus for dialog assistants. The dataset contains 20.6 hours of face-to-face acted dialogs recorded in a professional studio (44.1 kHz stereo) and segmented into 11,796 utterances across 3 speakers. Unlike read-speech resources, Dialogs captures turn-taking rhythm and expressive prosody, and provides per-utterance style/emotion labels spanning 12 categories. We validate corpus quality with crowd MOS tests, showing comparable audio quality and intelligibility to strong Russian studio baselines while achieving higher ratings for expressiveness and conversational naturalness. Finally, we train a VITS2 model as a proof of concept, demonstrating that Dialogs supports training expressive, dialog-like TTS despite limited per-speaker data. 
\end{abstract}

\section{Introduction}

Modern conversational assistants and text-to-speech (TTS) systems increasingly demand natural, expressive, and interactive speech. Achieving such naturalness depends critically on training and adaptation data: high-quality recordings that capture prosodic variation, natural timing, interruptions and other features of real conversation are essential for producing realistic dialog behaviour in TTS systems.

\subsection{Gap analysis}

For Russian, there is a notable shortage of studio-quality conversational corpora suitable for fine-tuning expressive TTS models. Existing Russian resources either lack expressive variation, consist of single-speaker read speech (Ruslan and Natasha each feature a single speaker), or are large but recorded under uncontrolled conditions. Table~\ref{tab:corpus_comparison} compares the most relevant existing corpora; high-quality English dialog datasets (Expresso\cite{Nguyen2023}, DailyTalk\cite{Lee2023}) are included as reference points. No existing Russian corpus combines studio-quality audio, conversational style, and per-utterance emotion/style annotations at this scale.

\begin{table}[t]
  \caption{Comparison with related speech corpora. Conv.\ = conversational/dialog recording.}
  \label{tab:corpus_comparison}
  \centering
  \small
  \setlength{\tabcolsep}{3pt}
  \begin{tabular}{lrrcccl}
    \toprule
    Corpus & Hrs & kHz & Styles & Conv. & Lang & License \\
    \midrule
    \textbf{Dialogs} (ours) & 20.6 & 44.1 & 12 & $\checkmark$ & RU & OpenRAIL \\
    Dusha \cite{Kondratenko2022} & 350 & 16 & 5 & -- & RU & Apache~2.0 \\
    Ruslan \cite{Gabdrakhmanov2019} & 31 & 44.1 & -- & -- & RU & -- \\
    Golos \cite{Karpov2021} & 1240 & 16 & -- & -- & RU & Apache~2.0 \\
    RESD \cite{Aniemore2023} & 3.5 & 16/44 & 7 & -- & RU & -- \\
    Natasha & 12 & 22 & -- & -- & RU & -- \\
    ESpeech \cite{Petrov2025} & 4500+ & 44 & -- & -- & RU & -- \\
    Expresso \cite{Nguyen2023} & 40 & 44.1 & 26 & $\checkmark$ & EN & CC-BY~4.0 \\
    DailyTalk \cite{Lee2023} & 20 & 44.1 & -- & $\checkmark$ & EN & CC-BY-SA~4.0 \\
    \bottomrule
  \end{tabular}
\end{table}

\subsection{Our contribution}

To address this gap, we introduce Dialogs, a studio-quality expressive conversational Russian speech corpus. Main \mbox{properties:}
\begin{itemize}
    \item Duration: 20.6 hours of dialog recordings.
    \item Recording protocol: Dialog (actors seated facing one another in studio); stereo recording.
    \item Annotation: transcripts and style/emotion labels (annotation procedure described below).
    \item Intended use: training and evaluation of modern TTS systems (we demonstrate utility by training VITS2~\cite{Kong2023} on the corpus).
    \item License: OpenRAIL (open; commercial use permitted).
\end{itemize}

Dialogs differs from conventional read corpora in that actors used prepared texts as prompts but were allowed to deviate and improvise expressive realizations in context, yielding natural intonation and rhythm relevant to conversational TTS. We demonstrate that VITS2~\cite{Kong2023} can be trained on Dialogs and produces noticeably expressive, conversational output. The corpus is publicly available at \url{https://huggingface.co/datasets/langswap/dialogs-ru-emotional-conversations} under the OpenRAIL license, which permits free use including commercial applications — unlike most existing Russian speech corpora.

\section{Related Work}
\subsection{Corpus taxonomy}
In practice, most public speech corpora fall into one of three categories. \textit{Read-speech} corpora (audiobooks, news) offer clean audio and accurate transcripts, but the recording style is monotone and neutral — actors read to microphones, not to each other. \textit{Web-mined} corpora (transcribed YouTube, podcasts) are naturally expressive and conversational, but transcript quality is bounded by ASR accuracy, and recording conditions are uncontrolled. \textit{Studio-recorded} corpora — the standard in the commercial TTS industry — are recorded by professional voice actors in acoustically treated rooms over extended sessions, yielding production-quality audio with accurate labels; emotions in this category are necessarily acted, which allows reliable labelling and style balance. Such corpora are rarely released publicly due to their commercial value.

Dialogs belongs to the third category. Existing open Russian studio corpora (Ruslan, Natasha) are single-speaker read-speech without emotion labels — sufficient for neutral TTS but not for expressive or dialog-oriented systems. Dialogs adds multi-speaker dialog recording style and per-utterance emotion annotation while maintaining the same studio quality, and releases all of this under the OpenRAIL license.

\subsection{Expressive TTS and training data}
Expressive and emotion-conditioned TTS has shown that the quality and style of training data directly determine a model's prosodic range \cite{Barakat2024, Tu2022, Zhang2025}. Conversational and dialog-style corpora are particularly valuable because they supply natural turn-taking, varied intonation, and emotional coloring that read-speech data cannot provide.

Dialogs is designed to fill exactly this data gap for Russian: it provides the dialog recording style, studio quality, and emotion labelling needed to train expressive TTS, properties absent from existing Russian corpora (see Table~\ref{tab:corpus_comparison}).

\section{The Dialogs Dataset}

\begin{figure}[t]
  \centering
  \includegraphics[width=\linewidth]{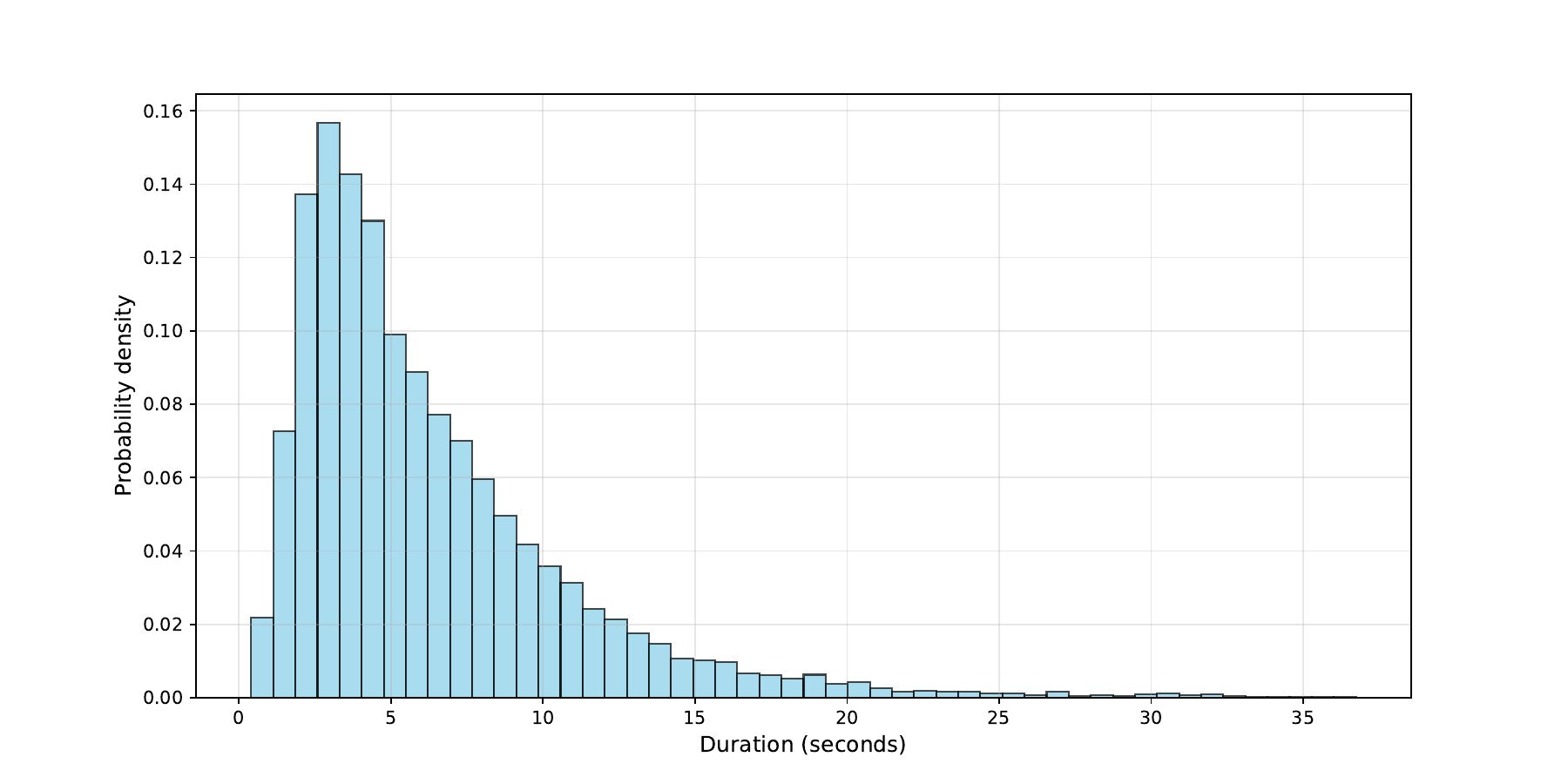}
  \caption{Audio recording duration distribution density.}
  \label{fig:audio_distribution}
\end{figure}

Dialogs contains 20.6 hours of expressive dialog speech recorded by professional puppet-theatre actors from Central Russia (see Figure~\ref{fig:audio_distribution} for the recording duration distribution; corpus statistics are given in Table~\ref{tab:corpus_stats}). Participant metadata, including performer experience and gender, are documented in the dataset metadata. All participants provided informed consent for recording and public release.

\begin{table}[t]
  \caption{Dialogs corpus statistics.}
  \label{tab:corpus_stats}
  \centering
  \begin{tabular}{ll}
    \toprule
    Property & Value \\
    \midrule
    Total duration & 20.6 h \\
    Speakers & 3 (M: 1, F: 2) \\
    \quad Masha (F) & 5{,}935 utt / 9.9 h \\
    \quad Sveta (F) & 3{,}616 utt / 4.4 h \\
    \quad Dima (M) & 2{,}245 utt / 6.2 h \\
    Utterances & 11{,}796 \\
    Avg.\ utterance duration & 6.3 s \\
    Vocabulary (unique words) & 21{,}262 \\
    Sample rate & 44.1 kHz, 16-bit \\
    Channels & Stereo \\
    Style categories & 12 \\
    Train / Dev / Test & 19.9 h / 0.30 h / 0.37 h \\
    UTMOS score (corpus) & 3.17 $\pm$ 0.07 \\
    License & OpenRAIL \\
    \bottomrule
  \end{tabular}
\end{table}

\subsection{Text material}
Recording scripts were created with the assistance of GPT-3.5 and cover a range of conversational scenarios: family interactions, travel, poetry recitation, tongue twisters, discussions about cars, and reviews of films and books, among others. Scripts served as prompts; actors were encouraged to use them as starting points and to improvise expressive variants as contextually appropriate.
To guide script generation, we defined a checklist of target textual features relevant for expressive Russian TTS:
\begin{itemize}
    \item Numbers and digits
    \item Addresses
    \item Neologisms / modern lexicon
    \item Improvised dialogues
    \item Paralinguistic events
    \item Read text
    \item Questions
    \item Exclamations
    \item Anglicisms i.e. brands and product names
    \item Variative: loud, quiet, fast, slow, screamed, and whispered
\end{itemize}

\subsection{Recording conditions}
Sessions were conducted in a professional studio. Actors sat facing each other. Recordings were captured in stereo using Behringer XM8500 microphones. Ambient noise level measured approximately 20 dBA. All files are provided in WAV format at 44.1 kHz, 16-bit.

\subsection{Preprocessing and segmentation}

Raw session recordings were manually segmented into utterances and aligned with transcripts; minimal noise reduction was applied. The corpus was split into train, development, and test subsets. The test set consists of the 188 utterances used in the corpus quality MOS evaluation (5 per speaker per emotion, stratified). The development set was constructed with the same stratified strategy from the remaining utterances (5 per speaker per emotion, random seed 42), yielding 180 utterances. All remaining utterances form the training set. Final split: 11{,}428 / 180 / 188 utterances (19.9~h / 0.30~h / 0.37~h).

\subsection{Annotation protocol}

Style/emotion annotation was performed on the Yandex Tasks crowd-sourcing platform. Each utterance is labeled with one or more style/emotion categories (neutral, happy, surprise, sad, disgust, angry, tongue-twister, poem, whisper, arrogance, laughing, fear) (see Table \ref{tab:styles_duration}). Each utterance was independently reviewed by 3 annotators; the majority label was selected as the ground truth, with ties (all three labels different) resolved by selecting the globally least-frequent category to encourage recall of rare styles.

\subsection{Corpus quality evaluation}

To verify recording quality and to directly compare Dialogs against the baseline corpora, we conducted crowd-sourced listening tests using the Yandex Tasks platform. For Dialogs, the evaluation subset was constructed by sampling 5 utterances per speaker per emotion category (188 utterances total, stratified across all 12 styles and 3 speakers; rare style--speaker combinations contributed fewer samples). For Ruslan and Natasha, which are single-speaker read-speech corpora without emotion labels, random subsets of 100 utterances each were drawn. The different subset sizes reflect the different corpus structures and do not affect comparability, as each condition's MOS estimate is reported with its own 95\% confidence interval. Native Russian listeners rated each clip on a 5-point scale across six dimensions: overall quality, audio quality, prosody, intelligibility, expressiveness, and conversational naturalness. Outlier annotators (identified by response pattern analysis) were removed prior to analysis, leaving 41 raters for Dialogs, 26 for Ruslan, and 23 for Natasha. Results are shown in Table~\ref{tab:corpus_mos}.

Dialogs matches Ruslan and Natasha on audio quality and intelligibility, confirming studio-level recording quality. Crucially, Dialogs scores substantially higher on expressiveness ($+0.23$–$0.25$ points) and conversational naturalness ($+0.24$–$0.30$ points) — the dimensions most relevant for dialog TTS applications.

\begin{table}[t]
  \caption{Corpus quality MOS by dimension (5-point scale, 95\% CI). Outlier annotators removed per corpus.}
  \label{tab:corpus_mos}
  \centering
  \small
  \setlength{\tabcolsep}{4pt}
  \begin{tabular}{lccc}
    \toprule
    \textbf{Dimension} & \textbf{Dialogs} & \textbf{Ruslan} & \textbf{Natasha} \\
    \midrule
    Overall          & $4.15 \pm 0.08$ & $4.26 \pm 0.09$ & $4.16 \pm 0.10$ \\
    Audio quality    & $4.19 \pm 0.08$ & $4.23 \pm 0.09$ & $4.18 \pm 0.11$ \\
    Prosody          & $\mathbf{4.05 \pm 0.08}$ & $4.01 \pm 0.10$ & $3.94 \pm 0.11$ \\
    Intelligibility  & $4.14 \pm 0.08$ & $4.17 \pm 0.09$ & $4.16 \pm 0.12$ \\
    Expressiveness   & $\mathbf{4.11 \pm 0.08}$ & $3.86 \pm 0.11$ & $3.88 \pm 0.13$ \\
    Conversational   & $\mathbf{4.08 \pm 0.08}$ & $3.78 \pm 0.13$ & $3.82 \pm 0.14$ \\
    \bottomrule
  \end{tabular}
\end{table}

\begin{table}[t]
  \caption{Total duration per style}
  \label{tab:styles_duration}
  \centering
  \begin{tabular}{ll}
    \toprule
    \textbf{Style name}      & \textbf{Duration (min)}                \\
    \midrule
neutral & 643.50 \\
happy & 349.59 \\
surprise & 78.59 \\
sad & 58.52 \\
laughing & 19.88 \\
angry & 19.19 \\
disgust & 17.40 \\
arrogance & 12.70 \\
tongue twisters & 12.13 \\
fear & 9.37 \\
poem & 8.82 \\
whisper & 6.13 \\

    \bottomrule
  \end{tabular}
\end{table}

\section{Experiments}

To demonstrate the utility of Dialogs as a training corpus, we train  VITS2~\cite{Kong2023}, a single-stage end-to-end TTS model on the Dialogs training split. The experiment serves as a proof-of-concept: the corpus spans three speakers with approximately 21 hours of audio total (9.9~h, 6.2~h, and 4.4~h per speaker respectively), a challenging regime for multi-speaker expressive TTS. We expect Dialogs to contribute expressiveness and conversational style when mixed with larger Russian corpora for production use. Training code is available at \url{https://github.com/shigabeev/vits2-emotional}.

The model was trained with a batch size of 16 on a single NVIDIA RTX 4090, following the original VITS2 hyperparameter defaults for 615000 steps.

We evaluate synthesized speech on a set of 20 held-out sentences per speaker, 60 audio files total, covering diverse contexts (questions, exclamations, numbers, emotional phrases), using MOS and UTMOS metrics:

\noindent\textbf{MOS.} Native Russian listeners rated synthesized speech across six dimensions — overall quality, audio quality, prosody, intelligibility, expressiveness, and conversational naturalness — on a 5-point scale via the Yandex Tasks crowd-sourcing platform, using the same protocol as the corpus quality evaluation.

\smallskip\noindent\textbf{UTMOS}~\cite{Saeki2022}\textbf{.} Automatic neural MOS predictor used as an additional objective naturalness metric. Higher is better.

\section{Results and Discussion}

Table~\ref{tab:result} reports evaluation scores for the VITS2 model trained on Dialogs. This experiment serves as a proof of concept that the corpus is sufficient to train a functional TTS system. Informal listening confirmed that synthesized speech exhibits clear conversational and expressive character consistent with the dialog recording style; the expressiveness ($2.56$) and conversational ($2.59$) dimension scores notably exceed intelligibility ($2.28$), showing that the model absorbs the prosodic style of the corpus. The MOS and UTMOS values are on the lower side due to the unbalanced per-speaker data budget (4.4--9.9 hours each); mixing Dialogs with a larger corpus is expected to yield production-quality results.

\begin{table}[b!]
  \caption{TTS evaluation results (VITS2 trained on Dialogs, 5-point scale, 95\% CI).}
  \label{tab:result}
  \centering
  \small
  \setlength{\tabcolsep}{4pt}
  \begin{tabular}{lc}
    \toprule
    \textbf{Dimension} & \textbf{VITS2-Dialogs} $\uparrow$ \\
    \midrule
    Overall          & $2.83 \pm 0.24$ \\
    Audio quality    & $2.97 \pm 0.23$ \\
    Prosody          & $2.55 \pm 0.21$ \\
    Intelligibility  & $2.28 \pm 0.20$ \\
    Expressiveness   & $2.56 \pm 0.21$ \\
    Conversational   & $2.59 \pm 0.21$ \\
    \midrule
    UTMOS (auto)~\cite{Saeki2022} & $3.36 \pm 0.06$ \\
    \bottomrule
  \end{tabular}
\end{table}

\subsection{Limitations} The corpus covers professional actors performing scripted prompts, so truly spontaneous dialog, overlapping speech, and background noise are absent. The training data is not balanced across speakers (4.4--9.9 hours each), which may affect per-speaker synthesis quality when the corpus is used in isolation; for production use, mixing with larger corpora is recommended.

\section{Generative AI Use Disclosure}
GPT-3.5 was used to assist in generating the recording scripts used for dataset collection. AI tools were also used heavily during development of evaluation pipelines (MOS templates, aggregation, table formation etc), collecting dataset statistics and for editing and polishing of this manuscript. All scientific content, experimental design and implementation are original and made by authors.

\bibliographystyle{IEEEtran}
\bibliography{export}

\end{document}